\documentclass[superscriptaddress,aps,preprintnumbers,amsmath,amssymb,prd,nofootinbib,preprint]{revtex4-1}
\pdfoutput=1

\usepackage{graphicx}
\usepackage{epstopdf}
\usepackage{dcolumn}
\usepackage{bm}
\usepackage{hyperref}
\usepackage{color}
\usepackage{setspace}

\begin{document}


\def\a{\alpha}
\def\b{\beta}
\def\c{\varepsilon}
\def\d{\delta}
\def\e{\epsilon}
\def\f{\phi}
\def\g{\gamma}
\def\h{\theta}
\def\k{\kappa}
\def\l{\lambda}
\def\m{\mu}
\def\n{\nu}
\def\p{\psi}
\def\q{\partial}
\def\r{\rho}
\def\s{\sigma}
\def\t{\tau}
\def\u{\upsilon}
\def\v{\varphi}
\def\w{\omega}
\def\x{\xi}
\def\y{\eta}
\def\z{\zeta}
\def\D{\Delta}
\def\G{\Gamma}
\def\H{\Theta}
\def\L{\Lambda}
\def\F{\Phi}
\def\P{\Psi}
\def\S{\Sigma}

\def\o{\over}
\def\beq{\begin{eqnarray}}
\def\eeq{\end{eqnarray}}
\newcommand{\gsim}{ \mathop{}_{\textstyle \sim}^{\textstyle >} }
\newcommand{\lsim}{ \mathop{}_{\textstyle \sim}^{\textstyle <} }
\newcommand{\vev}[1]{ \left\langle {#1} \right\rangle }
\newcommand{\bra}[1]{ \langle {#1} | }
\newcommand{\ket}[1]{ | {#1} \rangle }
\newcommand{\EV}{ {\rm eV} }
\newcommand{\KEV}{ {\rm keV} }
\newcommand{\MEV}{ {\rm MeV} }
\newcommand{\GEV}{ {\rm GeV} }
\newcommand{\TEV}{ {\rm TeV} }
\newcommand{\1}{\mbox{1}\hspace{-0.25em}\mbox{l}}
\newcommand{\headline}[1]{\noindent{\bf #1}}
\def\diag{\mathop{\rm diag}\nolimits}
\def\Spin{\mathop{\rm Spin}}
\def\SO{\mathop{\rm SO}}
\def\O{\mathop{\rm O}}
\def\SU{\mathop{\rm SU}}
\def\U{\mathop{\rm U}}
\def\Sp{\mathop{\rm Sp}}
\def\SL{\mathop{\rm SL}}
\def\tr{\mathop{\rm tr}}
\def\mpl{M_{PL}}

\def\IJMP{Int.~J.~Mod.~Phys. }
\def\MPL{Mod.~Phys.~Lett. }
\def\NP{Nucl.~Phys. }
\def\PL{Phys.~Lett. }
\def\PR{Phys.~Rev. }
\def\PRL{Phys.~Rev.~Lett. }
\def\PTP{Prog.~Theor.~Phys. }
\def\ZP{Z.~Phys. }

\def\dd{\mathrm{d}}
\def\ff{\mathrm{f}}
\def\BH{{\rm BH}}
\def\inf{{\rm inf}}
\def\ev{{\rm evap}}
\def\eq{{\rm eq}}
\def\SM{{\rm sm}}
\def\Mpl{M_{\rm Pl}}
\def\GeV{{\rm GeV}}
\newcommand{\Red}[1]{\textcolor{red}{#1}}

%


\title{Revisiting Scalar Quark Hidden Sector 
\\ in Light of $750$-GeV Diphoton Resonance}
\author{Cheng-Wei Chiang}
\affiliation{Center for Mathematics and Theoretical Physics and Department of Physics, National Central University, Taoyuan, Taiwan 32001, R.O.C.}
\affiliation{Institute of Physics, Academia Sinica, Taipei, Taiwan 11529, R.O.C.}
\affiliation{Physics Division, National Center for Theoretical Sciences, Hsinchu, Taiwan 30013, R.O.C.}
\author{Masahiro Ibe}
\affiliation{Kavli IPMU (WPI), UTIAS, University of Tokyo, Kashiwa, Chiba 277-8583, Japan}
\affiliation{ICRR, University of Tokyo, Kashiwa, Chiba 277-8582, Japan}
\author{Tsutomu T. Yanagida}
\affiliation{Kavli IPMU (WPI), UTIAS, University of Tokyo, Kashiwa, Chiba 277-8583, Japan}

\begin{abstract}
We revisit the model of a $CP$-even singlet scalar resonance proposed in 
\href{http://arxiv.org/abs/arXiv:1507.02483}{arXiv:1507.02483}, where the resonance appears as the lightest composite state made of scalar quarks participating in hidden strong dynamics.
We show that the model can consistently explain the excess of diphoton events with an invariant mass 
around $750$~GeV reported by both the ATLAS and CMS experiments.
We also discuss the nature of the charged composite states in the TeV range which accompany to the neutral scalar.
Due to inseparability of the dynamical scale and the mass of the resonance, the model also predicts signatures associated with the hidden dynamics such as leptons, jets along with multiple photons at future collider experiments.
We also associate the TeV-scale dynamics behind the resonance with an explanation of dark matter.
\end{abstract}

\date{\today}
\maketitle
\preprint{IPMU15-0225}

\section*{Introduction}

Recently, both the ATLAS and CMS Collaborations 
reported intriguing excess events in the search for a high-mass resonance 
decaying into diphotons in 13-TeV $pp$ collisions~\cite{ATLAS,CMS:2015dxe}.
The excess peaks at the diphoton invariant mass around $750$~GeV,
with significances being $3.6\,\s$ and $2.6\,\s$ 
by using  $3.2$~fb$^{-1}$ and   $2.6$~fb$^{-1}$ of data, respectively.
Using the model of a narrow scalar resonance, these local significances are 
reproduced when its production cross section times the branching ratio into diphotons are 
\begin{eqnarray}
\sigma(pp\to S) \times Br(S\to \g\g) &=& 6.0^{+2.4}_{-2.0}\,{\rm fb}\ , ({\rm ATLAS})\ ,\\
\sigma(pp\to S) \times Br(S\to \g\g) &=& 5.6^{+2.4}_{-2.4}\,{\rm fb}\ , ({\rm CMS})\ ,
\end{eqnarray}
respectively~\cite{Buttazzo:2015txu}.
(See also Refs.~\cite{Knapen:2015dap,Franceschini:2015kwy,McDermott:2015sck,Ellis:2015oso,Low:2015qep,Gupta:2015zzs,Dutta:2015wqh,Falkowski:2015swt,Alves:2015jgx,Kim:2015ksf,Berthier:2015vbb,Craig:2015lra} 
for phenomenological analyses of the resonance.)

After the reports, a plethora of models have been discussed to account for the signals.
Among them, models of (pseudo) scalar resonances originating from hidden strong 
dynamics have gathered particular attention, with its production at the LHC and decay 
into photons being explained via the gauge interactions of the constituents of the singlet composite 
state\,\cite{Harigaya:2015ezk,Nakai:2015ptz,DiChiara:2015vdm,Bellazzini:2015nxw,Matsuzaki:2015che,Bian:2015kjt,Liao:2015tow,Cline:2015msi,Belyaev:2015hgo}.
In this paper, we want to point out that an existing model proposed in Ref.~\cite{Chiang:2015lqa}
can consistently account for the diphoton signal while evading constraints from other high-mass resonance
searches made at the $8$-TeV LHC.
This model was originally proposed to explain the excess 
at around $2$~TeV in the searches for a diboson resonance in the ATLAS experiment~\cite{Aad:2015owa}.
As we will see, we can readily explain the $750$-GeV resonance by lowering 
the dynamical scale and mass parameters in the model.

In this model, the scalar resonance appears as the lightest composite state 
under hidden strong dynamics at around the TeV scale.
A peculiar feature of the model is that the hidden sector consists of scalar quarks,
and the lightest composite state is not pseudo-Goldstone bosons.
With this feature, the mass of the resonance should be in close proximity to the dynamical scale,
unlike in the models where the resonance is identified with a pseudo Goldstone modes.
As a result, the model predicts intriguing signatures associated with the hidden dynamics 
at the LHC such as leptons, jets and leptons with multiple photons as well as the 
existence of charged composite resonances in companion with the $750$-GeV resonance.
We also associate the TeV-scale dynamics behind the resonance with an explanation of dark matter.

\section*{A Scalar Resonance from Hidden Dynamics}

In the model of Ref.~\cite{Chiang:2015lqa}, the scalar resonance, $S$ with mass $M_S$, 
couples to the gauge bosons in the Standard Model (SM)  due to the SM gauge charges 
of the constituent hidden scalar quarks. 
Such interactions are described by the effective Lagrangian (see also Ref.~\cite{Barbieri:2010nc} for earlier works):
\begin{eqnarray}
\label{eq:SEFT}
{\cal L}_{\rm eff} =
 \frac{1 }{\L_{3}}S G^a_{\mu\nu}G^{a\,\mu\nu}
 + 
  \frac{1}{\L_2}S W^i_{\mu\nu}W^{i\,\mu\nu}
 + 
\frac{5}{3} \frac{1 }{\L_1}S B_{\mu\nu}B^{\mu\nu}\ ,
\end{eqnarray}
where, $\Lambda_{1,2,3}$ are suppression scales which are related to the dynamical scale of the hidden sector, and 
$G$, $W$ and $B$ are the field strengths of the $SU(3)_C$, $SU(2)_L$, and $U(1)_Y$ gauge bosons, 
respectively. 
These gauge fields are normalized so that their kinetic terms are given by
\begin{eqnarray}
{\cal L}=- \frac{1}{4g_s^2}  G^a_{\mu\nu}G^{a\,\mu\nu} 
- \frac{1}{4g^2} W^i_{\mu\nu}W^{i\,\mu\nu} 
-\frac{1}{4g^{'2}} B_{\mu\nu}B^{\mu\nu}\ ,
\end{eqnarray}
with $g_s$, $g$ and $g'$ being the corresponding gauge coupling constants, and the superscripts $a$ and $i$ denoting the indices for the corresponding adjoint representations.

Through the effective interaction with the gluons and in the narrow width approximation, the scalar resonance is produced at the LHC via the gluon fusion process 
\begin{eqnarray}
\sigma(p+p\to S) &=&
\frac{\pi^2}{8} \left(\frac{\Gamma(S\to g+g)}{M_S}\right) \times 
\left[\frac{1}{s}\frac{\q{\cal L}_{gg}}{\q\cal \t}
\right]\ ,\cr
\frac{\q{\cal L}_{gg}}{\q\cal \t} &=& \int_0dx_1 dx_2 f_g(x_1) f_g(x_2) \delta (x_1 x_2 - \tau)\ ,
\end{eqnarray}
where $\tau = M_S^2/s$ and $\sqrt{s}$ denotes the center-of-mass energy of the proton-proton collision.
Using the of MSTW2008 parton distribution functions (PDF's)~\cite{Martin:2009iq}, we obtain  
\begin{align}
\frac{1}{s}\frac{\q{\cal L}_{gg}}{\q\cal \t} &\simeq
\left\{\begin{array}{ll}
0.97\times 10^3~{\rm pb} & ({\rm for } \sqrt{s} = 8~{\rm TeV})\ ,\\
4.4 \times 10^3~{\rm pb} & ({\rm for } \sqrt{s} = 13~{\rm TeV})\ ,
\end{array}\right.
\end{align}
where we fixed the factorization scale and the renormalization scale at $\mu = M_S/2$ for $M_S \simeq 750$~TeV.%
\footnote{See {\it e.g.}, Ref.~\cite{Baglio:2010ae} for a discussion on higher-order QCD corrections, {\it i.e.}, the $K$-factor, for the production of the scalar resonance.}

The partial decay widths of the scalar resonance are given by
\begin{eqnarray}
\label{eq:SGG}
\G(S\to g+g) &=& \frac{2}{\pi} \left( \frac{g_s^2}{\Lambda_3}  \right)^2 M_S^3\ , \\
\label{eq:SWW}
\G(S\to W^+ + W^-) &=& \frac{1}{2} 
 \frac{1}{\pi} \left( \frac{g^2}{\Lambda_2} \right)^2 M_S^3\ , \\
 \label{eq:SZZ}
 \G(S\to Z+Z) &=& \frac{1}{4} 
 \frac{1}{\pi} \left[
 \left( \frac{g^2}{\Lambda_2}\right)  c_W^2
+
\frac{5}{3}\left(
\frac{g^{\prime 2}}{\Lambda_1} 
\right)s_W^2 
 \right]^2 M_S^3\ , 
 \\
 \G(S\to \g+\g) &=& \frac{1}{4} 
 \frac{1}{\pi} \left[ 
  \left( \frac{g^2}{\Lambda_2}\right)  s_W^2
+
\frac{5}{3}\left(
\frac{g^{\prime 2}}{\Lambda_1} 
\right)
c_W^2
 \right]^2 M_S^3\ , \\
 \G(S\to Z+\g) &=& \frac{1}{2} 
 \frac{1}{\pi} \left[ 
  \left( \frac{g^2}{\Lambda_2}\right)  
-
\frac{5}{3}\left(
\frac{g^{\prime 2}}{\Lambda_1} 
\right)
 \right]^2
 c_W^2s_W^2
 M_S^3\ , 
\end{eqnarray}
where $s_W \equiv \sin\theta_W$ and $c_W = (1-s_W^2)^{1/2}$ with $\theta_W$ being the weak mixing angle. 
The masses of the $W$ and $Z$ bosons are neglected to a good approximation.

Now let us discuss the model content and hidden dynamics that lead to the scalar resonance.
Following Ref.~\cite{Chiang:2015lqa}, we consider a set of scalar fields $Q$'s that carry both the hidden $SU(N_h)$ and the SM gauge charges.
The $SU(N_h)$ interaction is assumed to become strong at a dynamical scale $\Lambda_{\rm dyn}$.
Explicitly, we take $N_h = 5$. 
The charge assignments of $Q$'s are given in Table~\ref{tab:Q}.
It should be noted that we assign the SM gauge charges to $Q$'s in such a way that they form an
anti-fundamental representation of the minimal $SU(5)$ grand unified theory (GUT).

\begin{table}[t]
\caption{\sl \small
Charge assignments of the {bi-fundamental} scalars under the hidden $SU(5)$ and the SM gauge symmetries.
The SM gauge charges of the $Q$'s are assigned so that they form an anti-fundamental
representation of $SU(5)_{\rm GUT}$.
}
\begin{center}
\begin{tabular}{cccccc}
\hline\hline
& ~$SU(5)$~ & ~$SU(3)_C$~ & ~$SU(2)_{L}$~ & ~$U(1)_Y$~
\\
\hline
$Q_L$& ${\mathbf 5}$& ${\mathbf 1}$& ${\mathbf 2}$ & $1/2$
\\
$Q_D$& ${\mathbf 5}$& $\bar{\mathbf 3}$& ${\mathbf 1}$ & $-1/3$
\\
\hline\hline
\end{tabular}
\end{center}
\label{tab:Q}
\end{table}

The bi-fundamental scalars are assumed to have masses,%
\footnote{The bi-fundamental scalars also possess quartic couplings although they
are not relevant for the later discussion.}
 $m_{D,L}$:
\begin{eqnarray}
\label{eq:mass}
{\cal L} \supset -m_D^2 Q_D^\dagger Q_D - m_L^2  Q_L^\dagger Q_L\ .
\end{eqnarray}
When the masses of the scalar quarks do not exceed $\L_{\rm dyn}$, 
the lightest composite state is expected to be a $CP$-even neutral 
composite scalar that is a mixture of $Q_L^\dagger Q_L$, $Q_D^\dagger Q_D$, 
and a $CP$-even glueball. 
It should be emphasized here that the lightest neutral scalar is expected to be 
lighter than the other SM-charged composite states due to mixing, since the 
charged scalar composite fields are not accompanied by mixing partners.
This situation should be compared with models with fermionic bi-fundamental representations 
where the lightest singlet appears as a Goldstone boson mode.
In this case, one of the neutral Goldstone bosons becomes heavier than the SM-charged Goldstone bosons due to the chiral anomaly of the hidden gauge interaction.
Thus, if we further take the mass parameter of the colored hidden quark, $M_D$, larger than 
$\L_{\rm dyn}$, no neutral Goldstone boson remains lighter than the SM-charged ones,
such as the $SU(2)_L$ triplet Goldstone bosons.
In our scalar quark model, on the other hand, we expect that the neutral scalar boson remains
lighter than the SM-charged composite bosons even if we take $m_D$ larger than $\L_{\rm dyn}$ due to the mixing with the glueball.
This feature may be important when we discuss the phenomenology of the charged composite states
(see discussions at the end of this section).

In our analysis, we are interested in how the singlet $S$ couples to the SM gauge bosons.
For this purpose, we parametrize the relative contributions of $[Q_L^\dagger Q_L]$ and $[Q_D^\dagger Q_D]$ by a mixing parameter $\theta_Q$:
\begin{eqnarray}
S \propto \cos\h_Q \times [Q_L^\dagger Q_L] +  \sin\h_Q\times  [Q_D^\dagger Q_D] \ .
\end{eqnarray}
For example, the $Q_D^\dagger Q_D$ content is expected to be suppressed for $m_{D}\gg m_L$.
A quantitative estimation of $\theta_Q$ is, however, difficult due to the non-perturbative nature of the strong interaction.  Hence we  take $\h_Q$ as a free parameter in the following analysis.%
\footnote{We assume that the hidden strong dynamics does not cause spontaneous breaking of the SM gauge symmetries, although $\vev{Q^\dagger Q}$ is expected to be non-vanishing.
In particular, the mass terms of $Q$'s lead to  linear terms of the SM singlet composite scalars, resulting in 
$\vev{Q^\dagger Q} \neq 0$.}
For $m_D \gtrsim \Lambda_{\rm dyn}$, the second contribution can be effectively regarded as the glueball contribution that couples to the gauge bosons through the $Q_D$-loop diagrams 
(see also Ref.~\cite{Novikov:1977dq}).

To match the scalar resonance in the effective field theory onto the composite states,  
we rely on the Naive Dimensional Analysis (NDA)~\cite{Cohen:1997rt,Luty:1997fk}, leading to 
\begin{eqnarray}
S \simeq 
\frac{4\pi}{\k\L_{\rm dyn}}\cos\h_Q \times [Q_L^\dagger Q_L] 
+ \frac{4\pi}{\k\L_{\rm dyn}}\sin\h_Q\times  [Q_D^\dagger Q_D] \ ,
\end{eqnarray}
with a canonical kinetic term. 
The parameter $\k$ represents ${\cal O}(1)$ uncertainties of the NDA.
Altogether, we obtain the effective interactions of $S$ to the SM gauge bosons as
\begin{eqnarray}
\label{eq:SEFT3}
{\cal L}_{\rm eff} &=&
\frac{\k}{4\pi\L_{\rm dyn}}\sin\h_Q
\,
S G^a_{\mu\nu}G^{a\,\mu\nu}
 + 
 \frac{\k }{4\pi\L_{\rm dyn}}\cos\h_Q\,
 SW^i_{\mu\nu}W^{i\,\mu\nu}
\nonumber \\
&& +
 \frac{2\k}{4\pi\L_{\rm dyn}}
\left(
\frac{\sin\h_Q}{3} 
+
 \frac{\cos\h_Q}{2}
\right)
S
B_{\mu\nu}B^{\mu\nu}\ .
\label{eq:GGGG2}
\end{eqnarray}
As a result, the coefficients in the effective interactions given in Eq.\,(\ref{eq:SEFT}) are given  by
\begin{eqnarray}
\label{eq:relation}
\frac{1}{\L_3} = \frac{\k\sin\h_Q}{4\pi \L_{\rm dyn}}\ , \,\,\,\,
\frac{1}{\L_2} = \frac{\k\cos\h_Q}{4\pi \L_{\rm dyn}}\ , \,\,\,\,
\frac{1}{\L_1} =  \frac{\k}{4\pi \L_{\rm dyn}}
 \frac{6}{5}
\left(
\frac{\sin\h_Q}{3} 
+
 \frac{\cos\h_Q}{2}
\right)\ , 
\end{eqnarray}
Therefore, the production rates and the branching ratios are determined by two parameters,
$\sin\h_Q$ and $\L_{\rm dyn}$, in this model.

\begin{figure}[t]
\begin{center}
  \includegraphics[width=.5\linewidth]{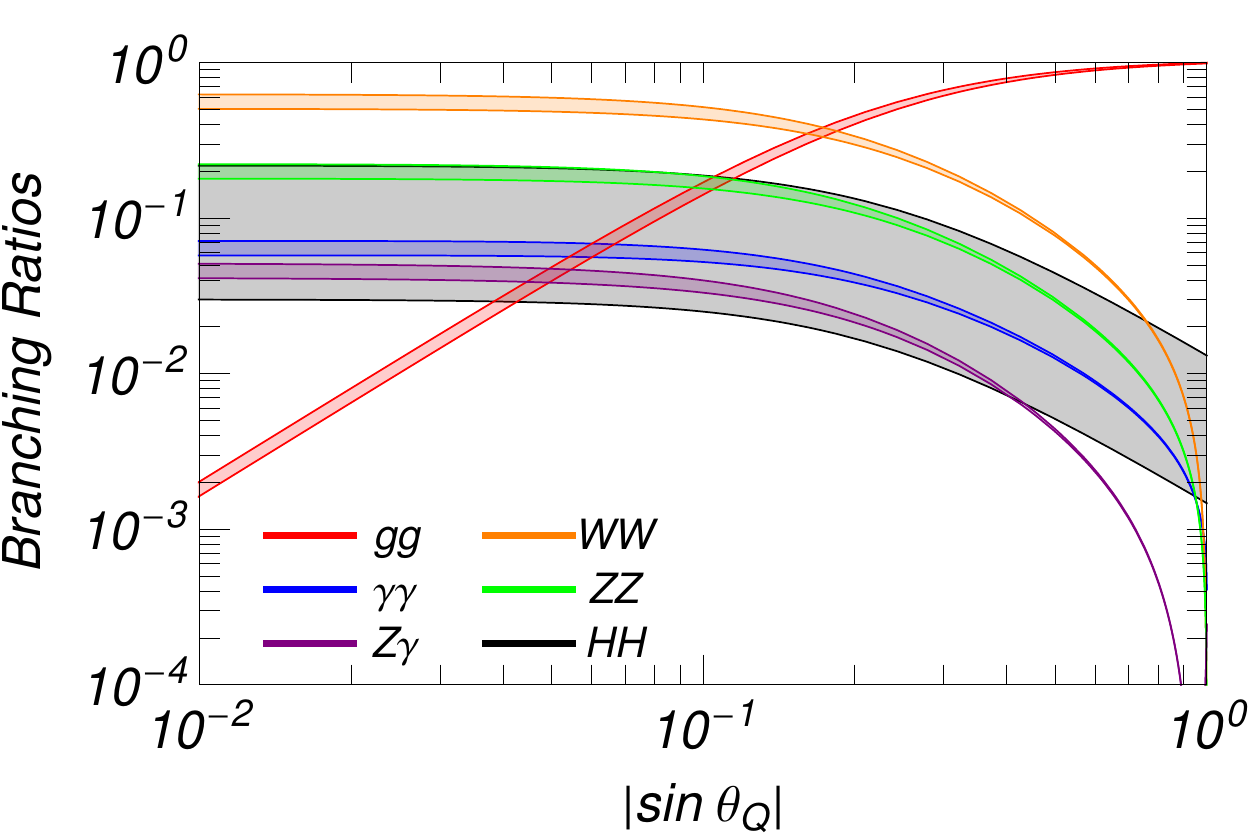}
  \end{center}
\caption{\sl \small
Branching ratios of the scalar resonance into the gauge boson pairs and the Higgs boson pair as functions of $\sin\theta_Q$ for $M_S = 750$~GeV and $\Lambda_{\rm dyn} = 1$~TeV.
The colored bands indicate the ranges of predictions as $\lambda$ is varied from $0.1$ to $0.3$.
}
\label{fig:branching}
\end{figure}

Fig.~\ref{fig:branching} shows the branching ratios of the scalar resonance as functions of $\sin\theta_Q$ for $M_S = 750$~GeV and $\Lambda_{\rm dyn} = 1$~TeV.
Here we use the running gauge coupling constants at the renormalization scale $M_S$.
The plot shows that the branching ratios of the $WW$ and $ZZ$ modes are about nine and three times larger than
that of the $\g\g$ modes for most of the parameter region.
On the other hand, the branching ratio into gluons is suppressed compared to even that of $\g\g$ for small $\sin\theta_Q$, as is evident from Eq.~(\ref{eq:relation}).

In the figure, we also take into account the decay of $S$ into a pair of the 125-GeV Higgs bosons due to interactions between $Q$'s and Higgs boson $H$,
\begin{eqnarray}
{\cal L} = \lambda_{L,D} \, Q_{L,D}^\dagger Q_{L,D} \, H^\dagger H \ ,
\end{eqnarray}
with $\l_{L,D}$ being coupling constants.
These interactions induce an effective interaction between $S$ and Higgs doublets,
\begin{eqnarray}
\label{eq:Higgs}
 {\cal L} =  \frac{\l}{4\pi} \L_{\rm dyn}S H^\dagger H\ , 
\end{eqnarray}
where we again use the NDA and reparameterize $\l_{L,D}$ and $\theta_Q$ by $\lambda$.
Through this operator, the resonance  decays into a pair of Higgs bosons with a partial decay width:%
\footnote{Strictly speaking, the operator in Eq.~(\ref{eq:Higgs}) also induces the decays into the weak gauge bosons.} 
\begin{eqnarray}
\G(S\to H+H^\dagger) = \frac{1}{8\pi M_S}\left( \frac{\lambda \L_{\rm dyn}}{4\pi}\right)^2 \ .
\end{eqnarray}
In Fig.~\ref{fig:branching}, we show the branching ratio of this mode for $\lambda = 0.1$--$0.3$.
We also show how the branching ratios into the gauge bosons are affected by 
the the Higgs pair mode as colored bands.
As is shown, the branching ratio into the Higgs bosons, proportional to $\lambda^2$, is subdominant for most of the parameter region.
Thus, its effects on the branching ratios of the modes of gauge boson pairs are not significant, as indicated by the narrow bands, and become diminishing when $\lambda$ is much smaller than $0.1$.

\begin{figure}[t]
\begin{center}
  \includegraphics[width=.5\linewidth]{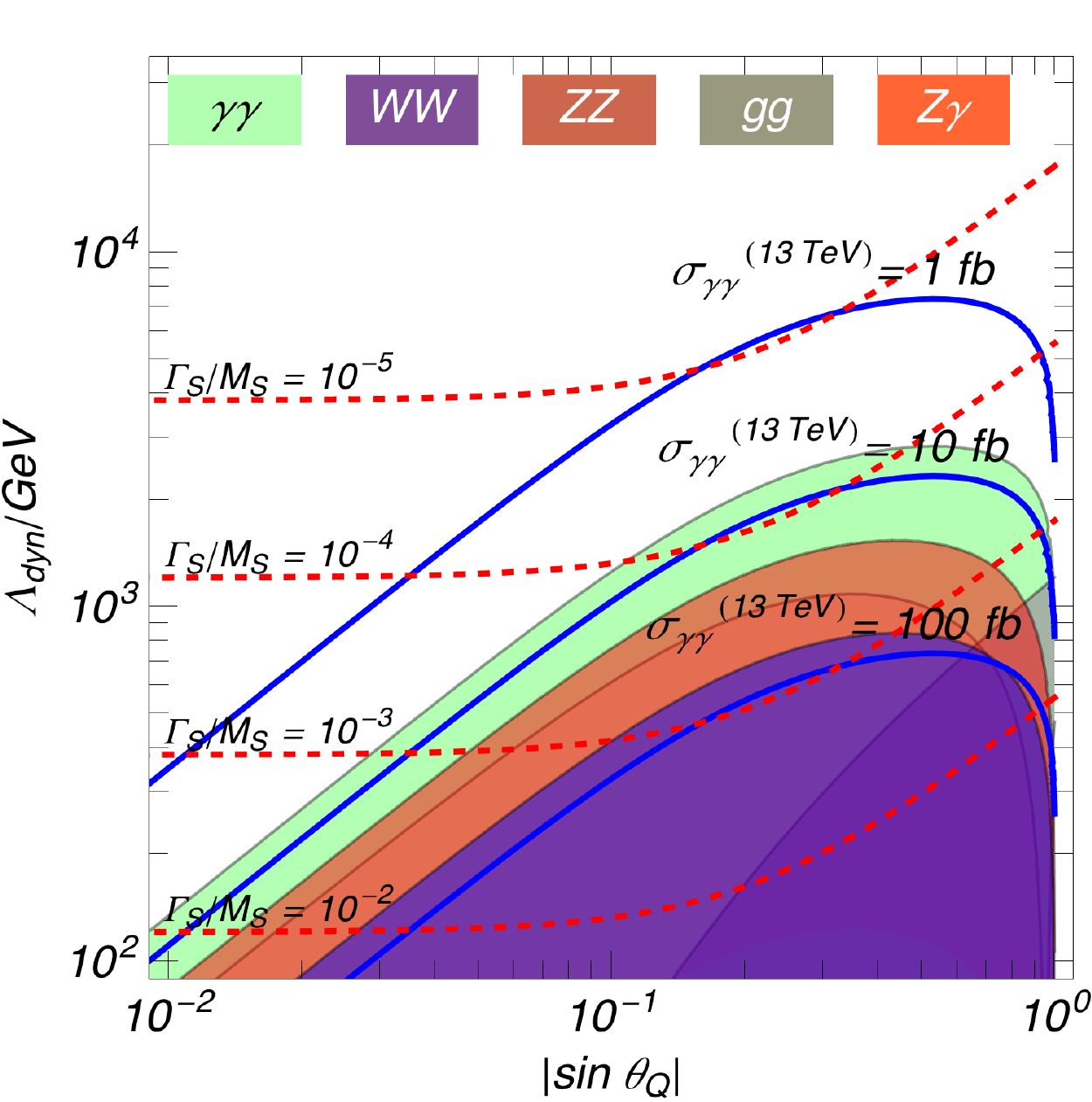}
  \end{center}
\caption{\sl \small
Contours of the production cross section of $S$ times its decay branching ratio into $\g\g$ at the $13$-TeV LHC by assuming the gluon fusion production process.
We fix $M_S = 750$~GeV and take the factorization and renormalization scales at $\mu = M_S/2$.
The color-shaded regions are excluded by the searches through various decay modes in LHC Run I 
as detailed in the main text.
The dashed red curves show the narrowness of the decay width.
}
\label{fig:crosssection}
\end{figure}

In the following, we discuss the preferred parameter region to explain the diphoton excess 
at $750$~GeV while being consistent with all the constraints from the searches for the other modes of gauge boson pairs.%
\footnote{The searches for a resonance decaying into a pair of Higgs bosons have 
imposed an upper bound of $\sigma(pp\to S \to hh) \lesssim 39$~fb~\cite{ATLAS:2014rxa}, 
which can be satisfied in most of the parameter region in Fig.~\ref{fig:crosssection}.}
In view of this, we simply neglect the effects of the operator in Eq.~(\ref{eq:Higgs}) by assuming $\lambda \lesssim 0.3$.%
\footnote{As discussed in Ref.~\cite{Chiang:2015lqa}, a similar quartic coupling between the 
$SU(2)_L$ composite triplet scalar and a pair of Higgs doublets leads to a non-vanishing vacuum expectation
value of the composite triplet scalar.  Due to electroweak precision constraints, the typical size of the quartic couplings should be at most ${\cal O}(0.1)$ for the model to be successful.}

Now, let us discuss the favored parameter region on the $(\sin\theta_Q, \Lambda_{\rm dyn})$ plane.
Fig.~\ref{fig:crosssection} shows in blue curves the contours of the cross section of the diphoton signal at 
the $13$-TeV LHC.
The colored regions are excluded by the resonance searches using various decays into gauge bosons
for $M_S=750$~GeV at the $8$-TeV LHC.
Here, we use the compilation of constraints listed in Ref.~\cite{Franceschini:2015kwy};
\begin{align}
\begin{split}
\sigma(pp\to S \to \g\g) &< 1.5\,{\rm fb}\quad  \mbox{\cite{CMS:2014onr, Aad:2015mna}} \ ,\\
\sigma(pp\to S \to WW) &< 40\,{\rm fb}\quad  \mbox{\cite{Khachatryan:2015cwa, Aad:2015agg}}\ , \\
\sigma(pp\to S \to ZZ) &< 12\,{\rm fb}\, \quad \mbox{\cite{Aad:2015kna}}\ , \\
\sigma(pp\to S \to Z\gamma) &< 4.0\, {\rm fb}\quad \mbox{\cite{Aad:2014fha}}\ , \\
\sigma(pp\to S \to jj) &< 2.5\,{\rm pb}\quad  \mbox{\cite{Aad:2014aqa,CMS:2015neg}} \ . 
\end{split}
\end{align}
The figure shows that the model can successfully explain the diphoton excess, $\s_{\g\g}^{(13\,{\rm TeV})} = {\cal O}(1)$\,fb
while evading all the above-mentioned constraints for  
\begin{eqnarray}
\L_{\rm dyn}  \simeq (10\,{\rm TeV}\mbox{--}30\,{\rm TeV}) \times \sin\h_Q\ .
\end{eqnarray}
Since the composite scalar mass is expected to be at around $\Lambda_{\rm dyn}$, 
we find that an appropriate range of the mixing angle is $\sin\h_Q  \simeq 10^{-1}\mbox{--}10^{-1.5}$.
This result also implies that the mass parameter $m_D$ is larger than $m_L$.  Moreover, $\Gamma_S/M_S \sim {\cal O}(10^{-4})$, justifying our narrow width approximation.  

Let us also comment on the production cross sections of the other gauge boson modes.
In the favored parameter region,  $\sin\h_Q  \simeq 10^{-1}\mbox{--}10^{-1.5}$, 
the branching ratios of the $WW$, $ZZ$, $Z\gamma$ modes are almost constant 
as a function of $\sin\h_Q$, while that of the $gg$ modes simply scales by $\sin^2\h_Q$.
Thus, the the production cross sections of the other gauge boson modes are predicted to be
\begin{eqnarray}
\s(pp\to S \to WW)&\simeq& 9\times \s(pp\to S \to \g\g)\ ,\\
\s(pp\to S \to ZZ)&\simeq& 3 \times \s(pp\to S \to \g\g)\ , \\
\s(pp\to S \to Z\gamma)&\simeq& 0.7 \times \s(pp\to S \to \g\g)\ , \\
\s(pp\to S \to jj)&\simeq& 270\,\sin^2\h_Q \times \s(pp\to S \to \g\g)\ ,
\end{eqnarray}
respectively, which will be tested by the LHC Run-II experiments.%
\footnote{If the coupling to the Higgs doublet in Eq.~(\ref{eq:Higgs}) is sizable, the above pretictions can be slightly modified.}

Before closing this section, let us comment on the SM-charged composite states 
predicted in this model.
Since the hidden sector consists of $Q_D$ and $Q_L$, the model predicts 
not only the singlet composite scalar, but also the charged composites:
an $SU(3)_C$ octet, an $SU(2)_L$ triplet,
and a bi-fundamental representation of $SU(3)_C\times SU(2)_L$ with 
a hypercharge of $5/6$.

Due to the color charge of the octet scalar, it is directly produced through the $SU(3)_C$ gauge interaction 
at the LHC and decays into a pair of gluons.
By the searches at the $8$-TeV LHC, the production cross section of the octet scalar with a mass around 
1\,TeV is constrained to be around ${\cal O}(1)$~pb~\cite{Aad:2014aqa,Khachatryan:2015sja},
which is much larger than the pair production cross section of the octet scalar \cite{GoncalvesNetto:2012nt}
as well as the single production rate via Eq.~(\ref{eq:SEFT}).
It should be noted that the octet scalar mass is expected to be larger than that of $S$
because $S$ is the lightest admixture of $[Q_L^\dagger Q_L]$ and $[Q_D^\dagger Q_D]$
while the octet is a unique scalar state.

Similarly, the triplet scalar is produced via the Drell-Yan process and immediately decays into SM electroweak gauge bosons and Higgs bosons through the interaction in Eq.~(\ref{eq:Higgs}).
Unlike the neutral scalar $S$, the triplet scalar does not couple to the gluons via dimension-$5$ operators.
To date, there is no stringent constraint on the triplet scalar with a mass of ${\cal O}(1)$\,TeV.

The scalar in the bi-fundamental representation of $SU(3)_C\times SU(2)_L$ requires special care, 
as it cannot decay into a pair of SM gauge bosons due to its charges.
To make it decay promptly, we introduce one flavor of fermions under the hidden $SU(5)$ gauge symmetry
 $(\psi_Q, \bar{\psi}_Q)$, which allow $Q$'s to couple to the SM quarks and leptons, $\bar{d}_R$  and $\ell_L$, via
\begin{eqnarray}
{\cal L} \supset y\, Q_D^\dagger\, \psi_Q \, \bar{d}_R + 
y\, Q_L^\dagger\, \psi_Q \, \ell_L + 
M_Q \psi_Q \bar{\psi}_Q + \mbox{h.c.}
\end{eqnarray}
Here, $y$ denotes a coupling constant and $M_Q$ the mass of the fermion $\psi_Q$.%
\footnote{We take $M_Q$ to be much larger than a TeV, so that they are not produced at the LHC.}
Through these interactions, the $[Q_D^\dagger Q_L]$ bound states decay 
into $\bar d_R^\dagger+\ell_L + S$~
\footnote{The two-body decay width into $\bar{d}_R^\dagger$ and $\ell_L$ is suppressed by 
the mass of the masses of the quarks.}
which is estimated to be roughly
\begin{eqnarray}
\Gamma \sim \left(\frac{1}{16\pi}\right)^2 \left(\frac{y^2}{4\pi}\right)^2\frac{\Lambda_{\rm dyn}^2}{M_Q^4} M_{[Q_D^\dagger Q_L]}^3\ ,
\end{eqnarray}
where $M_{[Q_D^\dagger Q_L]} $ denotes the mass of the bound state.
For $M_Q \lesssim 10^4$\,GeV, the bound state decays promptly into down-type quarks and leptons
and $S$ which subsequently decays into jets, $WW$, $ZZ$, $Z\gamma$ or $\gamma\gamma$ as discussed before.

For a larger $M_Q$, {\it e.g.}, $M_Q \gtrsim 10^7$~GeV, the bound state can be stable within the detectors and give an striking signature.
The lower mass limit put by the results of searches for heavy stable charged particles at CMS
ranges up to $0.9$--$1$~TeV, depending on the QED charges\,\cite{Chatrchyan:2013oca}.%
\footnote{When the mass of the scalar $[Q_D^\dagger Q_L]$ bound state is $1$~TeV,
the production cross section is $0.2$~fb at $8$~TeV~\cite{Kramer:2012bx} 
and $6$~fb at $13$ TeV~\cite{Borschensky:2014cia}.}
It should also be noted that for $M_Q \gg 10^{8}-10^9$~GeV, the lifetime of the bound state 
becomes longer than ${\cal O}(1)$ second and spoils the success of the Big Bang Nucleosynthesis~\cite{kkm}.%
\footnote{See also Ref.~\cite{Hamaguchi:2007rb} for related discussions.}

\section*{Discussions and Conclusions}

In this paper, we have revisited a model of scalar composite resonance that 
couples to the SM gauge bosons via the higher-dimensional operators proposed in
Ref.~\cite{Chiang:2015lqa} in light of the $750$~GeV diphoton excess discovered recently in the LHC Run-II.
In this model, the lightest composite state is expected to be the $CP$-even singlet 
scalar which is the admixture of the neutral bi-linear composite of 
the scalar quarks and a glueball. 
As we have shown, the model can consistently explain the excess while 
evading all the constraints from other high-mass resonance
searches made in LHC Run-I.
It should be noted that the $CP$ property of the resonance 
can be tested by measuring the angular distribution of the four leptons in the final sates 
of the $ZZ$ modes (see, {\it e.g.}, Ref.~\cite{Godbole:2007cn}).
Thus, this composite scenario can be clearly distinguished from the other composite models
where the neutral scalar manifests as a $CP$-odd pseudo-Goldstone mode.

The neutral scalar boson is accompanied by many charged bound states whose masses are also in the TeV regime.
Therefore, we expect that the LHC Run-II experiments will discover a zoo of such particles around that scale.
In particular, the bound state of $[Q_L^\dagger Q_D]$ has a striking signature of decaying into 
a lepton, a down-type quark and $S$, or it can even leave charged tracks inside the detector 
when the bound state is sufficiently stable.

As a peculiar feature of this model, 
the mass of the lightest composite state is not separable from the dynamical scale of the hidden sector, 
as it is not protected for any symmetry reasons.
Thus, the dynamical scale should be in close proximity to the composite mass, unlike again the models
in which the $750$-GeV resonance is identified as a pseudo-Goldstone boson.
Therefore, we expect that the quark-like picture of the hidden sector emerges at a rather low energy
in future collider experiments.
For example, production of multiple partons in the hidden sector becomes possible and ends up 
with events of multiple jets, multiple leptons and multiple photons.

Before closing this paper, let us address an important question: ``who ordered the $750$-GeV resonance?"
One ambitious answer is the dark matter candidate.
In fact, as discussed in Ref.~\cite{Chiang:2015lqa}, this model has a good dark matter candidate:
the lightest baryonic scalar
\begin{eqnarray}
B \propto  QQQQQ\ .
\end{eqnarray}
This state is neutral under the SM gauge groups due to the choice of $N_h = 5$.%
\footnote{To make $B$ stable, a (discrete) symmetry is required.
We presume that such a symmetry is not broken spontaneously by the strong dynamics as long 
as $m_{D,L}^2 \neq 0$.}
It should be emphasized that the neutralness of the lightest baryonic state under the SM gauge group
 is one of the prominent features of this model.
If, instead, the hidden sector consists of bi-fundamental fermions, the neutral baryonic state
is expected to be heavier than the lightest but SM-charged baryonic state since 
the neutral baryonic state has a larger orbital angular momentum inside.

In the early universe, the baryonic scalars annihilate into a pair of lighter scalar nnon-baryonic composite states. 
The thermal relic abundance would be much lower than the observed dark matter density if the annihilation cross section (into $S$, glueballs, etc.) saturates the unitarity limit~\cite{Griest:1989wd}. 
The abundance of $B$ is roughly given by,
\begin{eqnarray}
\Omega_B h^2 \sim 0.25\times 10^{-3} \frac{1}{F(M_B)^4}\left( \frac{M_B}{5\,{\rm TeV}}\right)^2 \ ,
\end{eqnarray}
where $M_B$ is the mass of $B$ and $F(M_B)$ denotes the form factor of the interactions of $B$
with the lighter states.%
\footnote{We define the form factor in such a way that the interaction of $B$ saturates the unitarity limit when $F = 1$.}
By remembering $M_B \gg \L_{\rm dyn}$ (in particular when $m_D > \L_{\rm dyn}$), 
it is expected that the form factor is slightly smaller than $1$.
Therefore, the thermal relic abundance of $B$ can be consistent with the observed dark matter density,
although a quantitative estimation is difficult due to our inability to estimate the form factor precisely.

Finally, let us comment on the direct detection of the dark matter candidate.
The coupling between $Q$'s and the Higgs doublet in Eq.~(\ref{eq:Higgs}) also leads to 
a direct coupling between the scalar dark matter and the Higgs doublet,
\begin{eqnarray}
{\cal L} = \lambda_B \,B^\dagger B H^\dagger H \ ,
\end{eqnarray}
where $\lambda _B $ is of ${\cal O}(\lambda_{L,D})$.%
\footnote{The dark matter also annihilates into Higgs bosons via this interactions, although 
its annihilation cross  section is subdominant for $\lambda_B = {\cal O}(0.1)$\,\cite{Kanemura:2010sh}. }
Thus, the dark matter interacts elastically with nuclei via the Higgs boson exchange
resulting in a spin-independent cross section~\cite{Kanemura:2010sh}
\begin{eqnarray}
\s_{SI}=\frac{\l_B^2}{4\pi m_h^4} \frac{m_N^4 f_N^2}{M_B^2} \simeq 5.4 \times 10^{-46}\,{\rm cm}^2
\times \l_B^2\left(\frac{5\,{\rm TeV}}{M_B}\right)^2\ ,
\end{eqnarray}
where we have used the lattice result $f_N\simeq 0.326$\,\cite{Young:2009zb}.
Although this coupling is much smaller than the current limit $\s_{SI} \lesssim 5 \times 10^{-44}$\,cm$^2\,(M_B/5\,{\rm TeV})$ by the LUX experiment\,\cite{Akerib:2015rjg}, it is within the reach of the proposed 
LUX-Zeplin (LZ) experiment~\cite{Akerib:2015cja}, with details depending on the coupling constants and the dark matter mass.

\section*{Acknowledgements}
This work is supported in part by the Ministry of Science and Technology of Taiwan under Grant No.~MOST104-2628-M-008-004-MY4 (C.-W.~C); Grants-in-Aid for Scientific Research from the Ministry of Education, Culture, Sports, Science, and Technology (MEXT), Japan, No. 24740151 and No. 25105011 (M.~I.) as well as No. 26104009 (T.~T.~Y.); 
Grant-in-Aid No. 26287039 (M.~I. and T.~T.~Y.) from the Japan Society for the Promotion of Science (JSPS); and by the World Premier International Research Center Initiative (WPI), MEXT, Japan (M.~I., and T.~T.~Y.).
MI is grateful to the Center for Mathematics and Theoretical Physics and Department of Physics 
of National Central University for hospitality where this work is completed.

\end{document}